\documentclass[aps,prd,superscriptaddress]{revtex4}
\usepackage{graphicx}
\usepackage{caption}
\usepackage{subfigure}
\usepackage{dcolumn}
\usepackage{setspace}
\usepackage{doi}
\usepackage{amsmath}
\usepackage{amssymb}
\usepackage{bm}
\usepackage{color}
\usepackage[normalem]{ulem}
\usepackage[dvipsnames]{xcolor}
\usepackage{hyperref}
\hypersetup{colorlinks=true, linkcolor=blue, citecolor=magenta}
\usepackage{autobreak}
\allowdisplaybreaks

\begin{document}

\title{Throttling process of Rotating Bardeen AdS Black Holes}
\author{Nan Chen}
\email{201721140006@mail.bnu.edu.cn} 
\affiliation{Department of Physics, Beijing Normal University, Beijing 100875, China}
\date{\today}

\begin{abstract}
	In this paper, the author study the throttling process of the Rotating Bardeen-AdS black hole in a systematic way. In the exended phase space,
	the mass of black holes should be viewed as entheply. We derive the Joule-Thomson coefficient $\mu$ explicitly, and with numerical method, we depict the inversion 
	curves and isenthalpic curves with different parameter $J$ and $g$. It is found that there are only minimum inversion temperature but no maximum inversion temperture, and the ratio between minimum inversion temperature $T_{min}$ and critical temperature $T_C$ is a little greater than $1/2$, and increase with the nonlinear parameter $g$. Furthermore, the shapes of the
	isenthalpic curves are similar to most case studied before, we calculate the inversion point exists as long as the mass is no less than a certain value $M_{min}$, the effect of the parameter $g$ on the throttling process is also discussed.
\end{abstract}

\maketitle

\section{INTRODUCTION}
Ever since the discovery of the striking resemblance between the machanics laws of black holes and the laws of thermodynamics\cite{Bardeen:1973gs,Bekenstein:1973ur}, also with the
demonstration of the Hawking radiation~\cite{Hawking:1974sw}, black holes are widely considered as a thermodynamic systems that possess its temperture and entropy,
thus the study of the thermodynamical properties of black holes has become a fascinating research area in theoretical physics for several decades.
The first remarkable work were done by Hawking and Page\cite{Hawking:1982dh}, who found that a first-order phase transition can occur in the Schwarzschild black hole in
the asymptotically Anti-de Sitter spacetime, this progress intrigued physicists to study the phase structure of black holes. What's more, motivated
by the profound discovery of AdS/CFT correspondence\cite{Maldacena:1997re}, many physicists had focused their attention on the study of the black holes thermodynamics
in the AdS spacetime. For example, the charge AdS black holes was found that share some analogous phase structures with the Van der Waals gas-fluids system\cite{Chamblin:1999tk,Chamblin:1999hg}.

Few years ago, some researchers proposed an innovative idea that the cosmological constant can be treated as a thermodynamical variable under some reasonalbe arguments\cite{Kastor:2009wy,Dolan:2010ha,Dolan:2011xt}.
In this view point, the cosmological constant is associated with the dynamical pressure, and its conjugate quantity is defined as the thermodynamic volume of the black hole.
In this frame work, the term $PdV$ is included in the first law of black holes, thus the phase space is extended, which is now known extended phase space, and the mass of AdS black hole is identified with the enthalpy rather than the internal energy. Based on this perspective, the phase structure in the charged AdS black hole system was reanalyzed in the extended phase space\cite{Kubiznak:2012wp}, the analogy
between the RN-AdS black hole and the VdW liquid-gas system had precisely established. This analogy has been generalized to many other AdS black holes, such as
the higher dimensional RN AdS black hole and Born-Infeld AdS black hole\cite{Gunasekaran:2012dq,Zou:2013owa}, the Gauss-Bonnet AdS black hole\cite{Cai:2013qga}, the Kerr-Newman AdS black hole\cite{Cheng:2016bpx}, and so forth.

Apart form the above research, recently, a well-known process in classical thermodynamics was generalized to AdS black holes, which is the throttling process, also
named as Joule-Thomson(JT) expandsion. A special feature of the throttling process is that the enphalpy remains unchanged during this process, in the circumstance of classical Van-de Waals gas, we can find in any general thermodynamics teaching material that there exist a minimum and a maximum inversion temperture, and the isenthalpic curves are seperated in two parts by the inversion curves, one part is the cooling region, while the other is the heating region, the intersection point is called inversion point, where the JT coefficient is zero. Back to the study of black holes, remind that the mass of
a black hole in the extend phase space is considered as enthaply, hence, it is interesting to investigate the throttling process in different AdS black holes, a first example
was done in the case of the charged AdS black hole\cite{Okcu:2016tgt}, the authors calculate the JT coefficient and the inversion temperture, then plot the inversion curves and isenthalpic curves to study its properties, some novel features different from the case of VdW gas were found. Similar studies had quickly extended
to other black holes, such as kerr AdS black hole\cite{Okcu:2017qgo}, kerr-Newman AdS black hole\cite{Zhao:2018kpz}, charged Gauss Bonnet AdS black hole\cite{Lan:2018nnp} and so on\cite{Mo:2018rgq,Mo:2018qkt,Ghaffarnejad:2018exz,Chabab:2018zix,Cisterna:2018jqg,AhmedRizwan:2019yxk,Yekta:2019wmt,Guo:2019gkr,Li:2019jcd,Pu:2019bxf,Rostami:2019ivr,Haldar:2018cks,Sadeghi:2020bon,Guo:2019pzq,Lan:2019kak,Nam:2020gud,K.:2020rzl}.

However, the above-mentioned studies mainly focus on the AdS black holes with singularity. The occurence of singularity is a big defect of Einstein's General Relativity,
as long as some fairly general assumptions are satisfied, then the curvature singularity resulting from gravitational collapse is inevitable\cite{Hawking:1973uf}. It is widely acknowledged that
these singularities do not exist in the real physical world, the General Relativity must be modified by quantum effect in the high energy and microscopic regime to become a completed theory, which is called
quantum gravity. Unfortunately, we are far from a well defined quantum gravity theory. Hence, this propelled physicists to turn their attention to the study of constructing
the singularity-free black holes in classical General Relativity, called regular black holes, with the price that the strong energy condition violated. In 1968, Bardeen first constructed a regular black hole\cite{bar}, with a nonlinear parameter $g$ to avoid the singularity, replaced by a replusive de-Sitter core.
Then until 2000, two researchers proved that the Bardeen black hole was an exact solution in a model of spacetime coupled to a nonlinear electrodynamics\cite{AyonBeato:2000zs}. As the regular
black hole has attracted more and more attention in recent years, regualr black holes of other types were found~\cite{AyonBeato:1999rg,Hayward:2005gi}, and the thermodynamic properties also studied in such cases. The throttling process of Bardeen-AdS black holes has been studied in Ref~\cite{Li:2019jcd,Pu:2019bxf}, with the absence of singularity, the thermodynamic properties were shown a slight difference compare to the ordinary cases. But it is more important to investigate the rotating case, because rotaion is a common phenomenon in universe, by applying the Newman-Janis algorithm to the Bardeen black hole one can get the rotating version\cite{Bambi:2013ufa}, which characterized by $g$ and $J$, our paper will focus on the study of the throttling process of rotating Bardeen black hole in the AdS spacetime, to seek the combined impact of $g$ and $J$, with the above consideration, our study will be meaningful to some extent.

The paper is organized as follows. In Sec.~\ref{review}, we give a brief review of the rotating Bardeen-AdS black hole, and show the thermodynamical quantities in the extended
phase space. Then, in Sec.~\ref{Bardeen}, we derive the explicit expression of JT coefficient, plot the inversion curves and isenthalpic curves with different values of $g$ and $J$, discuss their effects, after that, the minimum inversion tenperature and critical temperature are computed, their ratio are shown, the minimum mass are also obtained. Finally, our conclusion and discussion will be demonstrated in the Sec.~\ref{con}.

\section{A BRIEF REVIEW OF ROTATING BARDEEN-ADS BLACK HOLE AND ITS THERMODYNAMIC QUANTITIES}\label{review}
The line element of the rotating Bardeen-AdS black hole can be found in Ref.\cite{Ali:2019myr}, which in Boyer-Lindquist coordinates reads,
\begin{equation}
\begin{aligned}
{{\it ds}}^{2}=&-{\frac {\Delta_{{r}}}{\Sigma} \left( {\it dt}-{\frac {{{\it a\,sin}}^{2}\theta}{\Xi}}d\phi \right) ^{2}}+{\frac {\Sigma\,}{\Delta_{r}}}{{\it dr}}^{2}+{\frac {\Sigma}{\Delta _{\theta }}}{d\theta }^{2}\\
&+{\frac {\Delta_{{\theta}}\,{\sin}^{2}\theta}{\Sigma} \left( {\it a\,dt}-{\frac { {a}^{2}+{r}^{2}}{\Xi}} d\phi  \right) ^{2}}\,,
\end{aligned}
\end{equation}
where
\begin{equation}
\Sigma =r^2+a^2\cos^2\theta\, ,\quad\quad\Xi =1-\frac{a^2}{l^2}\,,
\end{equation}
\begin{equation}
\begin{aligned}
\Delta _r&=\left(a^2+r^2\right) \left(\frac{r^2}{l^2}+1\right)-2\tilde{m}(r)\,r\,,\\
\Delta _{\theta }&=1-\frac{a^2}{l^2}\cos ^2\theta\,,
\end{aligned}
\end{equation}
where $l$ is the curvature radius which is related to the cosmological constant via $\Lambda =-3/l^2$, in the extended phase space, we have the thermodynamic pressure $P=-\Lambda/8=3/8l^2$,
and the conjugated thermodynamic volume is defined by $V=({\partial M}/{\partial P})_{J,S,g}$. In Bardeen black hole, the term $\tilde{m}(r)$ is equal to $M\left(\frac{r^2}{r^2+g^2}\right)^{3/2}$, the parameter $g$ charecterize the nonlinear electrodynamics effect, while $\tilde{m}(r)=M-Q^2/2r$ in the KN black hole. One can verify the nonsingularity of this rotating Bardden-AdS black hole by computing its Riemannian tensor and scalar, it can be found that the curvature is finite over the spacetime as long as $g\neq0$. By treating the cosmological constant related to the thermodynamic pressure, the first law of black hole and the Smarr relation can be rewitten in the extended phase space
in the following form 
\begin{equation}
\begin{aligned}
dM&=TdS+{\varOmega}dJ+{\varPhi}dg+VdP\,,\\
M&=2TS+2{\varOmega}J+{\varPhi}g-2VP\,,\label{smarr}
\end{aligned}
\end{equation}
the mass $M$ can be regarded as a function(Eq.\ref{M}) of angular momentum $J$, entropy $S$ and the nonlinear parameter $g$, which is derived in Ref.\cite{Ali:2019myr}, in that literature, the temperature, entropy and other thermodynamic quantities are view as a function of the horizon radius $r_+$, in this paper, we can derive these quantities by taking the partial derivative of $M$ as following Eq.~\ref{T}-\ref{phi},
\begin{widetext}
	\begin{eqnarray}\label{M}
	M=\frac{{m}}{{ \Xi \mathop{{}}\nolimits^{{2}}}}
	=\frac{{\sqrt{{12 \pi \mathop{{}}\nolimits^{{2}}J\mathop{{}}\nolimits^{{2}}S{ \left( {3+8PS} \right) }+{ \left( {g\mathop{{}}\nolimits^{{2}} \pi +S} \right) }\mathop{{}}\nolimits^{{3}}{ \left( {3+8PS} \right) }\mathop{{}}\nolimits^{{2}}}}}}{{6\sqrt{{ \pi }}S}}\,,
	\end{eqnarray}
	
	\begin{eqnarray}\label{T}
	T=\left(\frac{{\partial M}}{{\partial S}}\right)_{J,P,g}=\frac{1}{8 \pi  M}\left(-\frac{2 \left(\pi  g^2+S\right)^3 \left(\frac{8 P S}{3}+1\right)}{S^3}+\frac{3 \left(\pi  g^2+S\right)^2 \left(\frac{8 P S}{3}+1\right)^2}{S^2}-\frac{4 \pi ^2 J^2}{S^2}\right)\,,
	\end{eqnarray}
	
	\begin{eqnarray}\label{V}
	{V={ \left( {\frac{{ \partial M}}{{ \partial P}}} \right) }\mathop{{}}\nolimits_{{S,J,P}}=\frac{{1}}{{9 \pi M}}{ \left[ {12 \pi \mathop{{}}\nolimits^{{2}}J\mathop{{}}\nolimits^{{2}}+\frac{{2}}{{S}} \left( g\mathop{{}}\nolimits^{{2}} \pi +S \left) \mathop{{}}\nolimits^{{3}}{ \left( {3+8PS} \right) }\right. \right. } \right] }}\,,
	\end{eqnarray}
	
	\begin{eqnarray}\label{omega}
	{ \Omega ={ \left( {\frac{{ \partial M}}{{ \partial J}}} \right) }\mathop{{}}\nolimits_{{S,P,g}}=\frac{{ \pi J}}{{MS}}{ \left( {1+\frac{{8PS}}{{3}}} \right) }}\,,
	\end{eqnarray}
	
	\begin{eqnarray}\label{phi}
	{ \Phi ={ \left( {\frac{{ \partial M}}{{ \partial g}}} \right) }\mathop{{}}\nolimits_{{S,P,J}}=\frac{{1}}{{12MS\mathop{{}}\nolimits^{{2}}}}{ \left[ {g{ \left( {3+8PS} \right) }\mathop{{}}\nolimits^{{2}}{ \left( {g\mathop{{}}\nolimits^{{2}} \pi +S} \right) }\mathop{{}}\nolimits^{{2}}} \right] }}\,,
	\end{eqnarray}
	
\end{widetext}
in the above expressions, only Eq.\ref{M}-\ref{V} will be used frequently in the following study, be careful of these quantities are function of $S$, $P$, $J$ and $g$, as one can easily observe that when $g=0$, all the above expressions will reduce to the case of Kerr-AdS black hole, which has been analyzed thoroughly in Ref.\cite{Caldarelli:1999xj}. 

\section{Throttling PROCESS OF THE ROTATING BARDEEN-ADS BLACK HOLE}\label{Bardeen}
In this section, we investigate the throttling process of the rotating Bardeen-AdS black hole in detail. The throttling process in classical thermodynamics is a gas with high pressure passes through a porous plug to another section with a low preesure in a thermally insulated tube and enthalpy remains constant during this process, the details can be found in any thermodynamics textbook. The change of temperture in the process is revealed by the Joule-Thomson coefficient, which is defined as
\begin{equation}\label{mu}
\begin{split}
\mu &=\left(\frac{\partial T}{\partial P}\right)_H   \\
&=\left(\frac{\partial T}{\partial P}\right)_M=\frac{1}{C_P}\left[T\left(\frac{\partial V}{\partial T}\right)_P-V\right],
\end{split}
\end{equation}
where $C_{P}=T(\partial S/\partial T)_P$ is the heat capacity at constant pressure. Due to the pressure always decreases during the process, as can be seen apparently, if the JT coefficient is positive, then the temperture decrease as the pressure decrease, which is called in the cooling region, on the contrary, while the JT coefficient is negative, the temperture increase as the pressure decrease, which is called the heating region. 

\subsection{JT coefficient}\label{JT}

As mentioned above, the JT coefficient is of primary significance, hence, we try to derive this quantity firstly. With Eq.\ref{T} and Eq.\ref{V}, it is straightforward to calculate the JT coefficient of the rotating Bardeen-AdS black hole as follows,
\begin{equation*}
\mu=A/B,
\end{equation*}
where
\begin{widetext}
	\begin{align}
	&A=\notag\\
	&4(-256 \pi ^6 g^{12} P^2 S^2-192 \pi ^6 g^{12} P S-36 \pi ^6 g^{12}-1344 \pi ^5 g^{10} P^2 S^3-1008 \pi ^5 g^{10} P S^2\notag\\
	&-189 \pi ^5 g^{10} S+512 \pi ^4 g^8 P^3 S^5-2304 \pi ^4 g^8 P^2 S^4-1944 \pi ^4 g^8 P S^3-378 \pi ^4 g^8 S^2-384 \pi ^5 g^6 J^2 P S^2\notag\\
	&-144 \pi ^5 g^6 J^2 S+2048 \pi ^3 g^6 P^3 S^6-896 \pi ^3 g^6 P^2 S^5-1536 \pi ^3 g^6 P S^4-342 \pi ^3 g^6 S^3-864 \pi ^4 g^4 J^2 P S^3\notag\\
	&-324 \pi ^4 g^4 J^2 S^2+3072 \pi ^2 g^4 P^3 S^7+1536 \pi ^2 g^4 P^2 S^6-144 \pi ^2 g^4 P S^5-108 \pi ^2 g^4 S^4-768 \pi ^3 g^2 J^2 P^2 S^5\notag\\
	&-1152 \pi ^3 g^2 J^2 P S^4-324 \pi ^3 g^2 J^2 S^3+2048 \pi  g^2 P^3 S^8+1728 \pi  g^2 P^2 S^7+432 \pi  g^2 P S^6+27 \pi  g^2 S^5-\notag\\
	&144 \pi ^4 J^4 S^2-768 \pi ^2 J^2 P^2 S^6-672 \pi ^2 J^2 P S^5-144 \pi ^2 J^2 S^4+512 P^3 S^9+512 P^2 S^8+168 P S^7+18 S^6),\\
	\notag\\
	&B=\notag\\
	&3 \sqrt{\pi } S(-16 \pi ^3 g^6 P S-6 \pi ^3 g^6+64 \pi ^2 g^4 P^2 S^3-9 \pi ^2 g^4 S+128 \pi  g^2 P^2 S^4+48 \pi  g^2 P S^3\notag\\
	&-12 \pi ^2 J^2 S+64 P^2 S^5+32 P S^4+3 S^3) \sqrt{(\pi  g^2+S)^3 (8 P S+3)^2+12 \pi ^2 J^2 S (8 P S+3)},
	\end{align}
\end{widetext}
the expression of the JT coefficient seems to be rather tedious, however, we can choose some specific values of $P,J$ and $g$ to analyze to behaviour of the JT coefficient as a fuction of entropy $S$.

\begin{figure*}[htbp]
	\subfigure[\quad $g=0.5$]
	{
		\centering
		\includegraphics[scale=0.57]{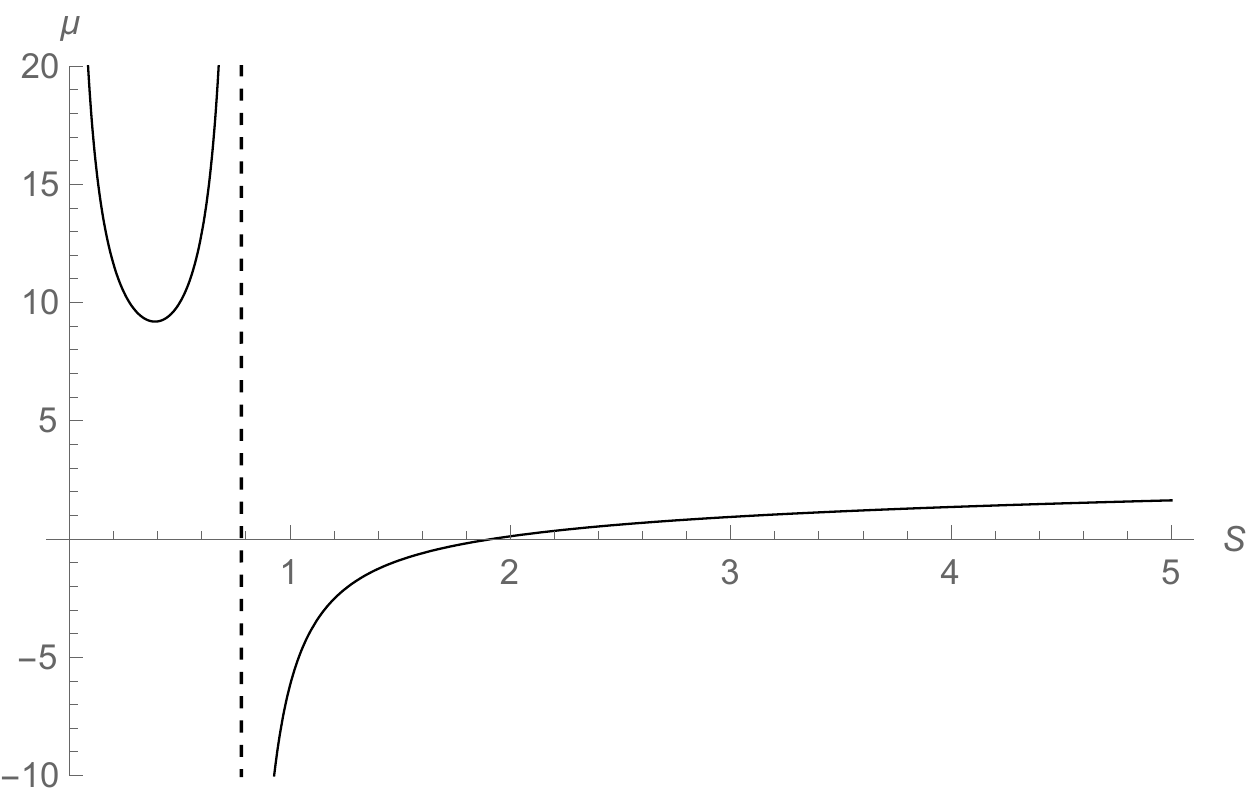}
	}
	\subfigure[\quad $g=1$]
	{
		\centering
		\includegraphics[scale=0.57]{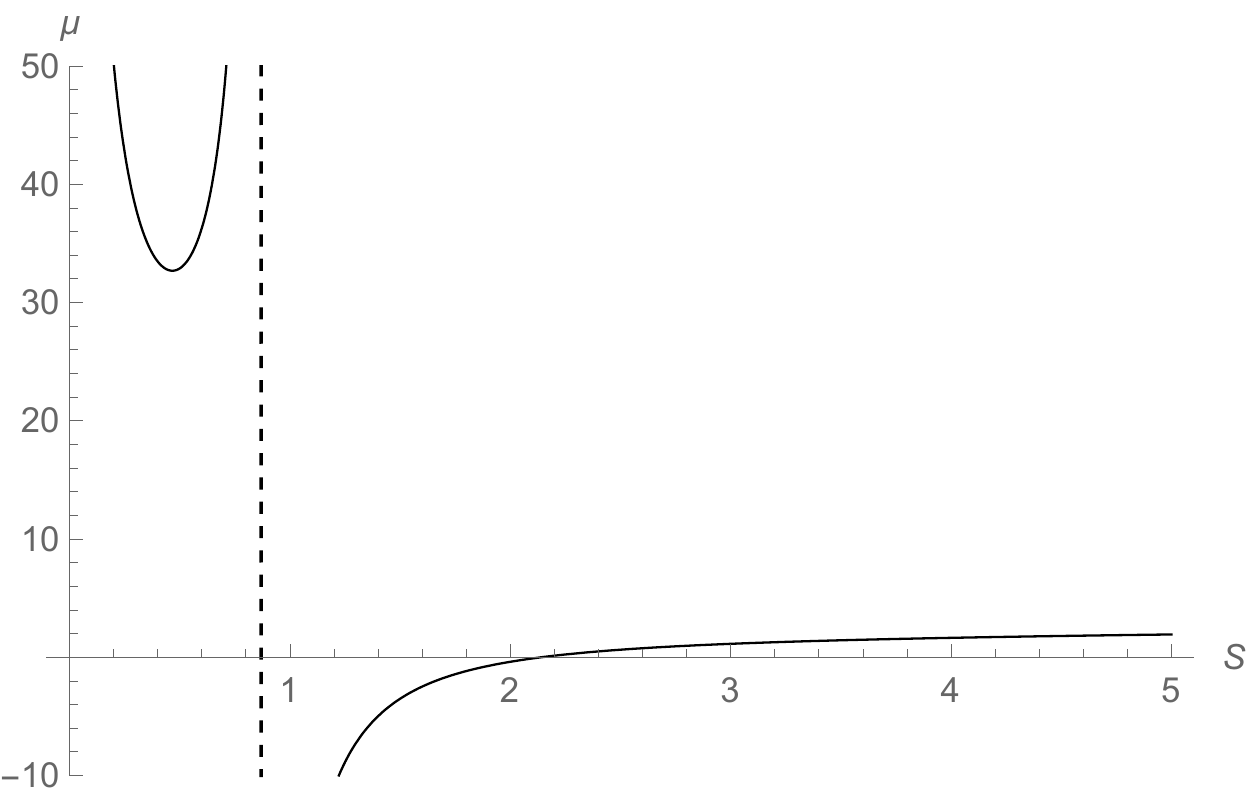}
	}
	\quad
	\subfigure[\quad $g=1.5$]{
		\centering
		\includegraphics[scale=0.57]{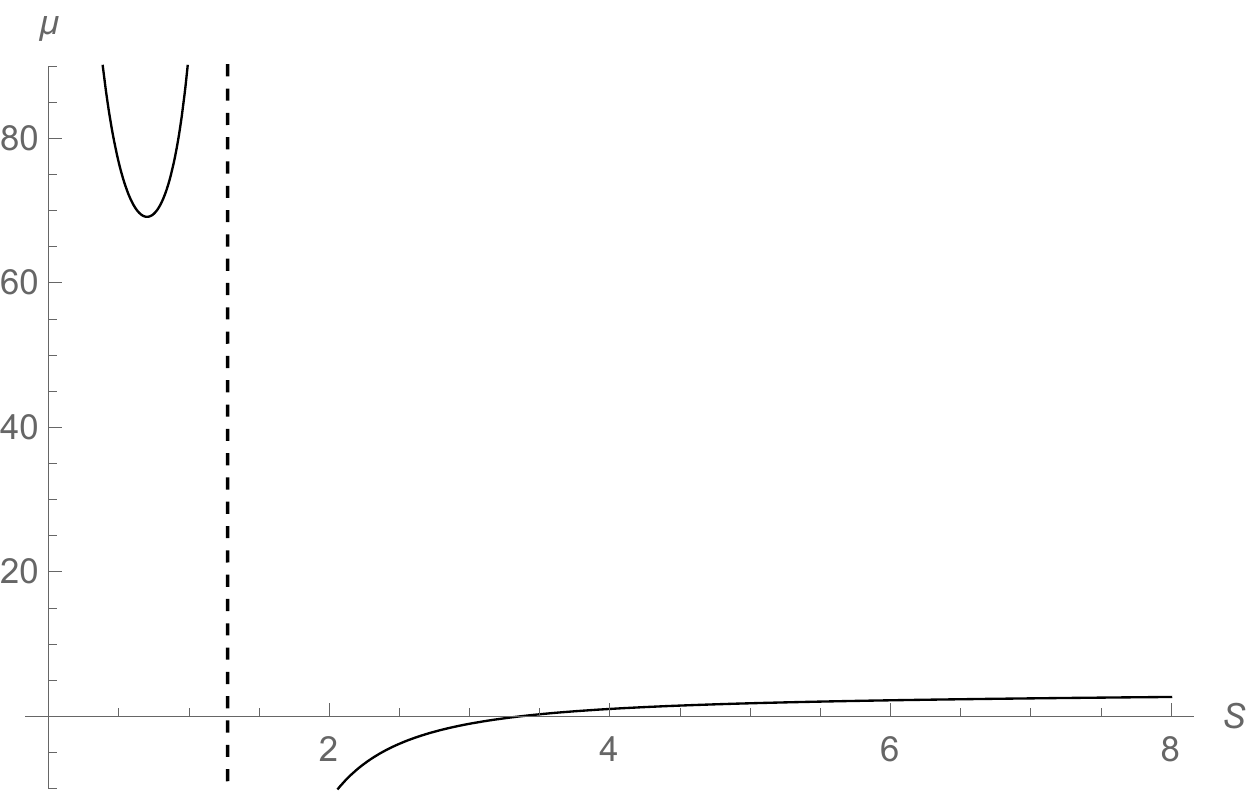}
	}
	\subfigure[\quad $g=2$]{
		\centering
		\includegraphics[scale=0.57]{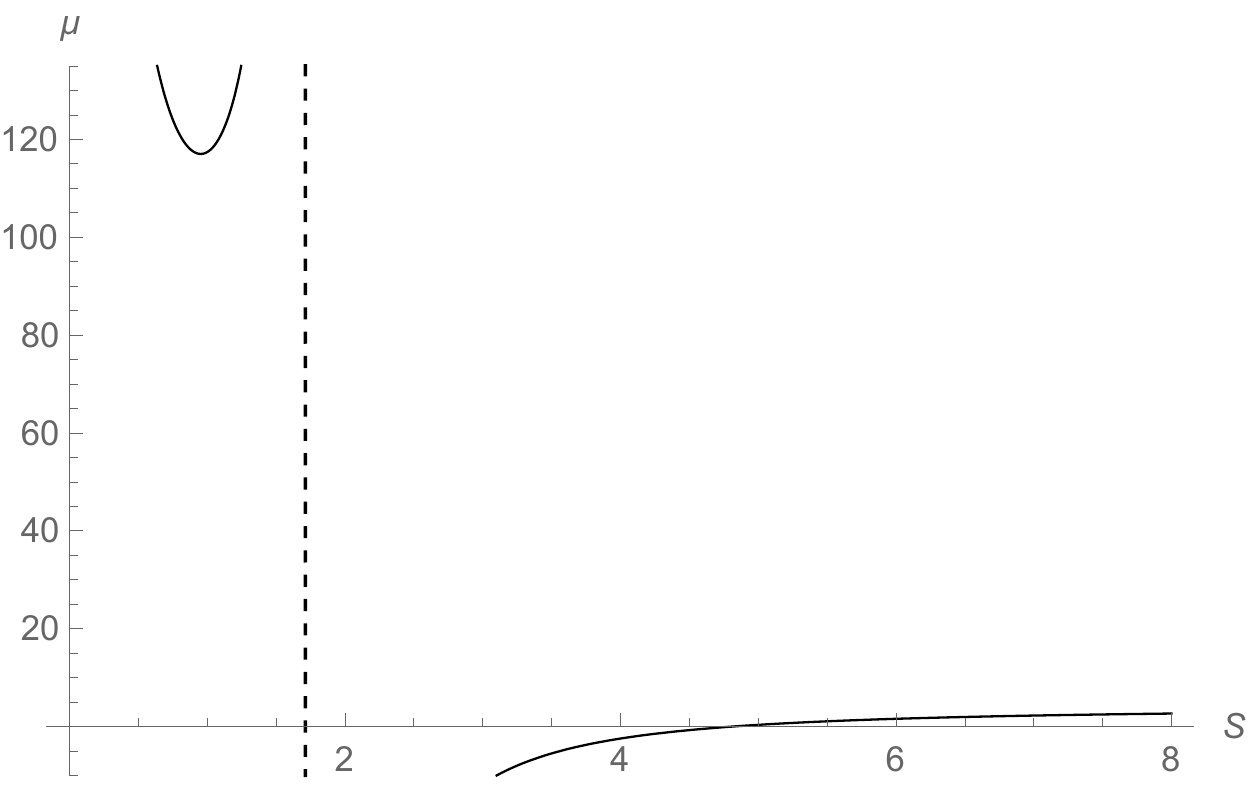}
	}
	\caption{JT-coefficient of the Rotating Bardeen-AdS black holes of different values of $g$ and with $J=1$, $P=1$.}
	\label{J-T}
\end{figure*}

 As we can see from Fig.~\ref{J-T}, there is a divergent point and zero point, and the total shape is similar to the case of Kerr-Newman AdS black hole\cite{Cheng:2016bpx}. Note that the divergent point of JT coefficient corresponds to the zero point of Hawking temperature. Moreover, it can be observed that the nonlinear parameter $g$ has a great infulence on the JT coefficient, as $g$ becomes larger, the divergent point and zero point of the JT coefficient are also become larger.

There is another way to calculate the JT coefficient which is first presented in Ref.\cite{Okcu:2017qgo}, we take the differential of the Smarr relation(Eq.\ref{smarr}) and use the conditions $dM=dJ=dg=0$ in the throttling process, we obtain,
\begin{equation}
SdT-PdV-2Vdp+Jd\Omega+\frac{g}{2}d\Phi=0,
\end{equation}
thus, the JT coefficient can also be expressed in the following form,
\begin{equation}
\begin{aligned}
\mu&=\left(\frac{\partial \,T}{\partial P}\right)_{M,J,g}\\
&=\frac{1}{S}\,\left[P\left(\frac{\partial \,V}{\partial P}\right)_{M,J,g}-J\left(\frac{\partial \,\Omega}{\partial P}\right)_{M,J,g}-\frac{g}{2}\left(\frac{\partial \,\Phi}{\partial P}\right)_{M,J,g}+2V\right],
\end{aligned}
\end{equation}

we can also calculate the JT coefficient in the above expression, but it is much more tedious because first we should derive the expression of $S$ from the Eq.\ref{M}, which maybe impossible, however, the final form must be consistent with each other.

\subsection{Inversion curves and Isenthalpic curves}\label{curve}

Next, with the JT coefficient, we may calculate the inversion temperature $T_i$ and then plot the inversion curves in the $T-P$ plane. In order to derive the $T_i$, we first set $\mu=0$, it is equivalent to set the numerator of $\mu$ to be zero, however, it does not lower the difficulty, the numerator of $\mu$ is an algebraic equation about $S$ of higher degree, it is out of the question to obtain the analytical form, we can only use numerical method to figure out the values of $S$ with different values of $P$, $J$ and $g$, and once we get the values of $S$, we subsititue it imto Eq.\ref{T} and the inversion temperature is directly obtained. 

\begin{figure*}[p]
	\subfigure[\quad $g=0.5,\,J=1,5,10,15$]
	{
		\centering
		\includegraphics[scale=0.57]{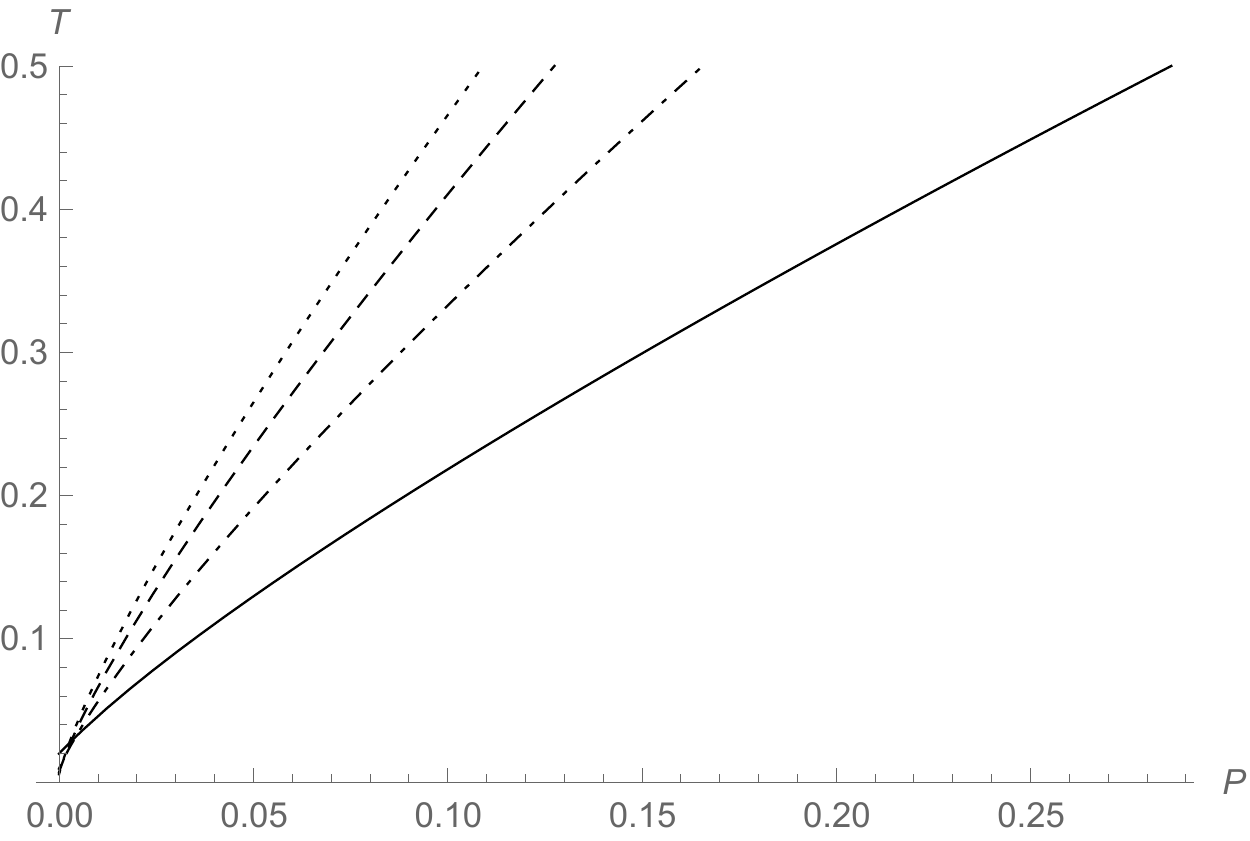}
	}
	\subfigure[\quad $J=1,\,g=0.5,1,1.5,2$]
	{\centering
		\includegraphics[scale=0.57]{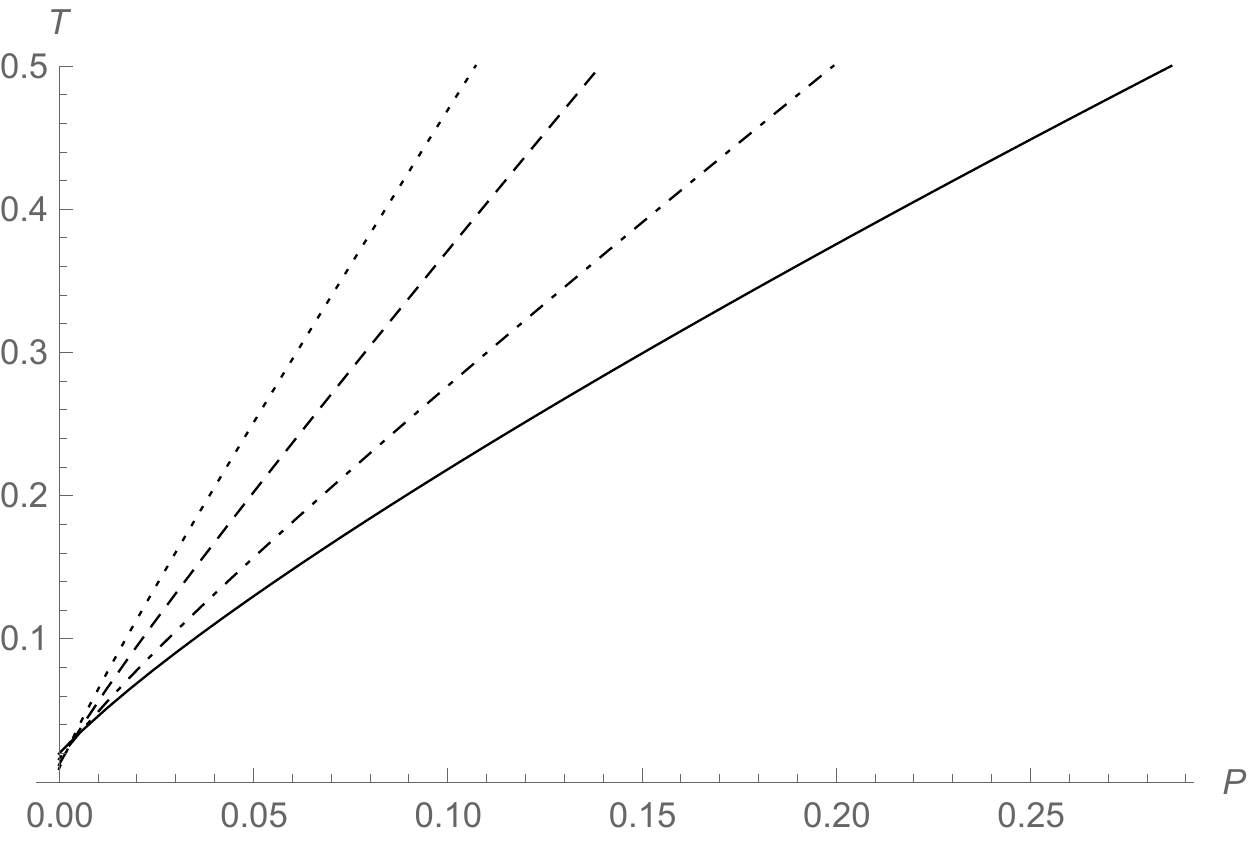}
	}
	\quad
	\subfigure[\quad $g=1,\,J=1,5,10,15$]
	{\centering
		\includegraphics[scale=0.57]{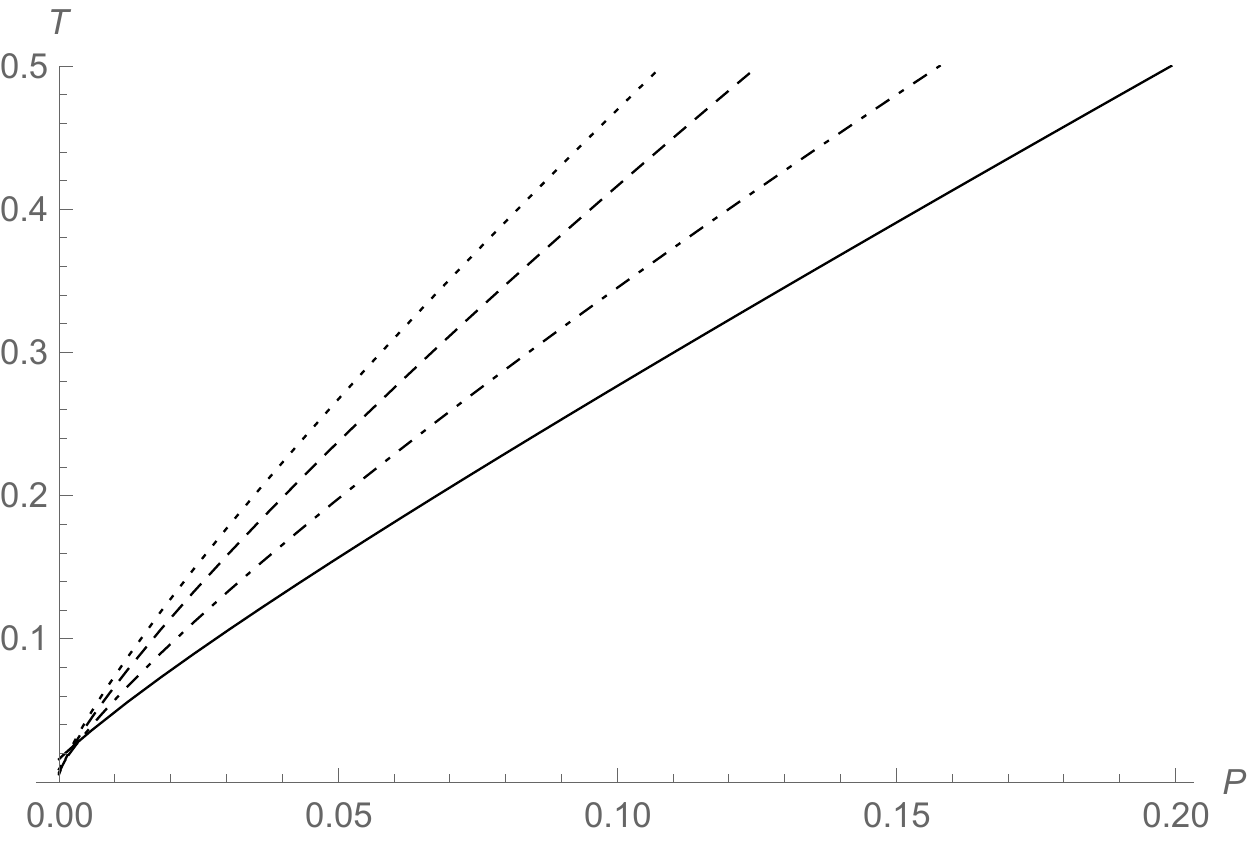}
	}
	\subfigure[\quad $j=5,\,g=0.5,1,1.5,2$]
	{\centering
		\includegraphics[scale=0.57]{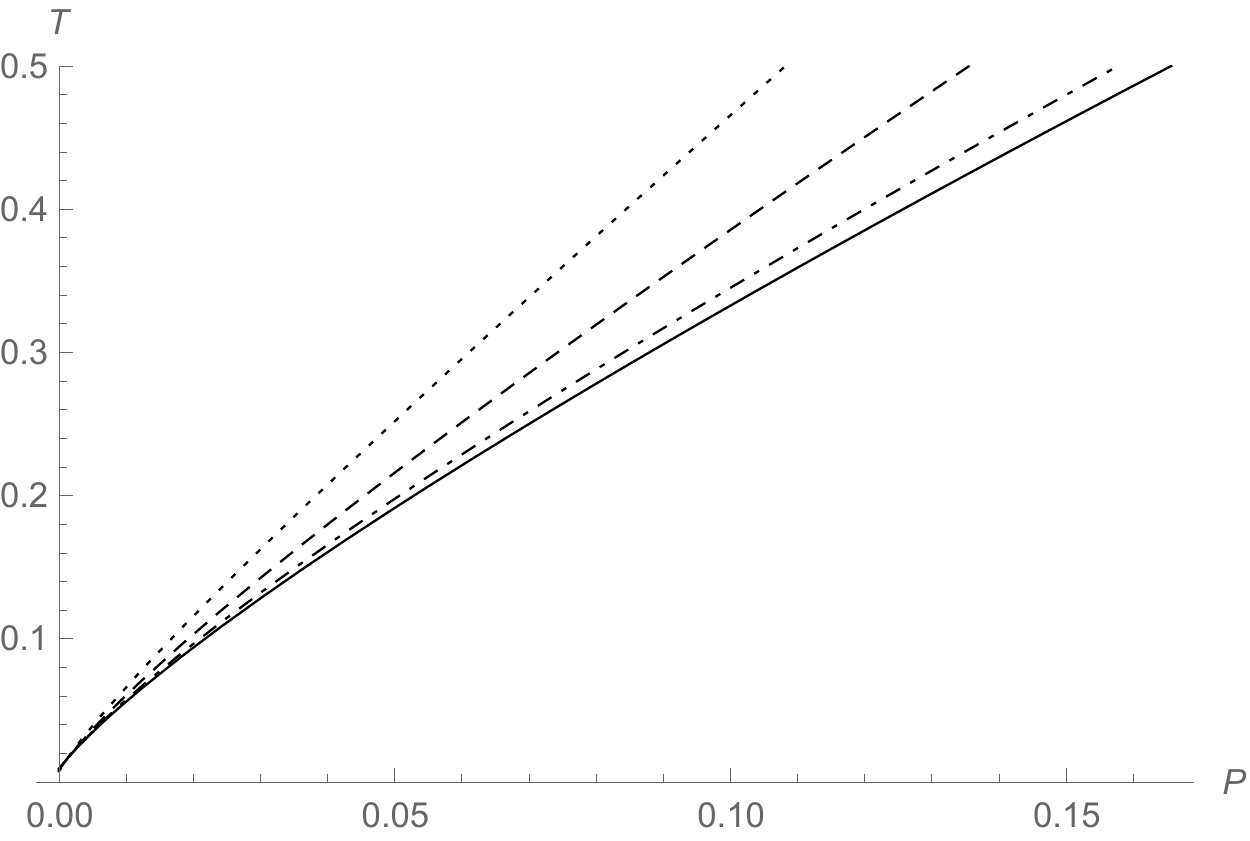}
	}
	\quad
	\subfigure[\quad $g=1.5,\,J=1,5,10,15$]
	{\centering
		\includegraphics[scale=0.57]{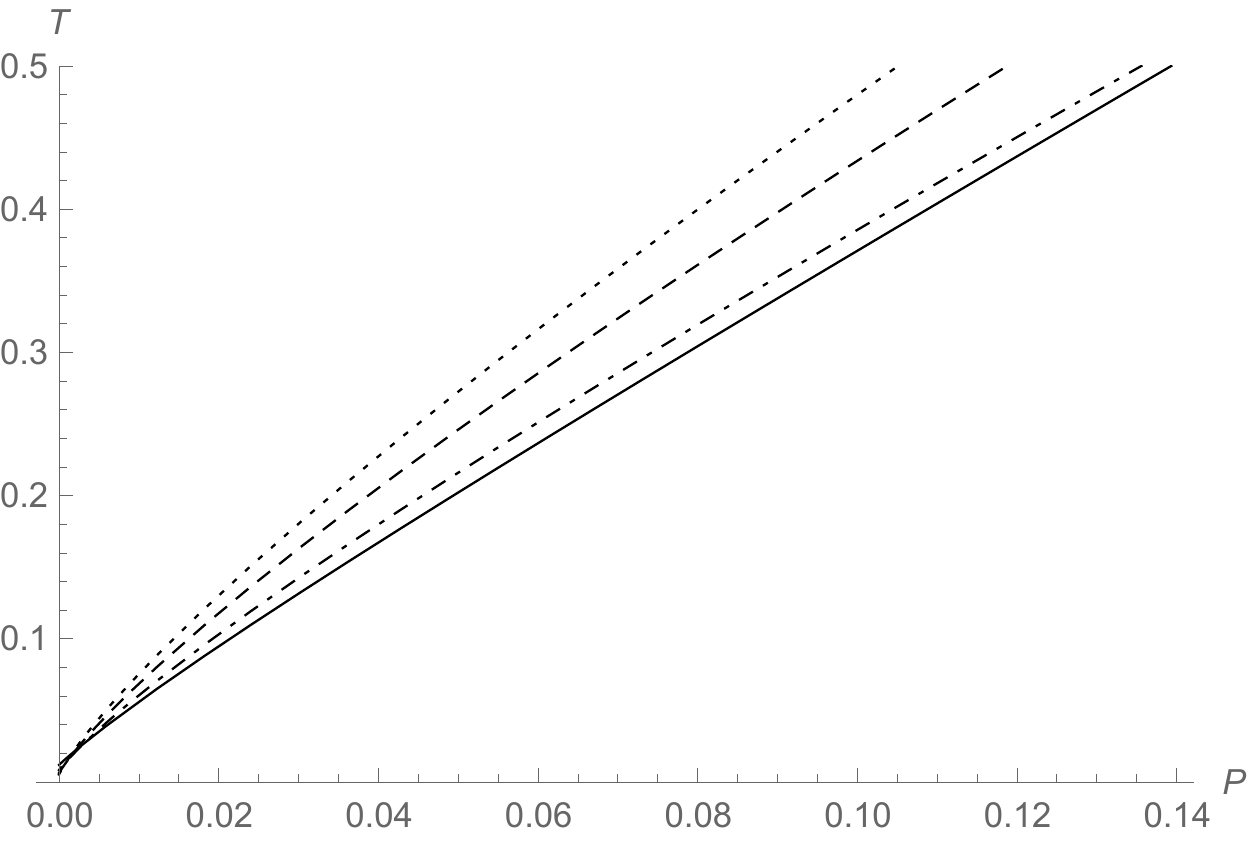}
	}
	\subfigure[\quad $j=10,\,g=0.5,1,1.5,2$]
	{\centering
		\includegraphics[scale=0.57]{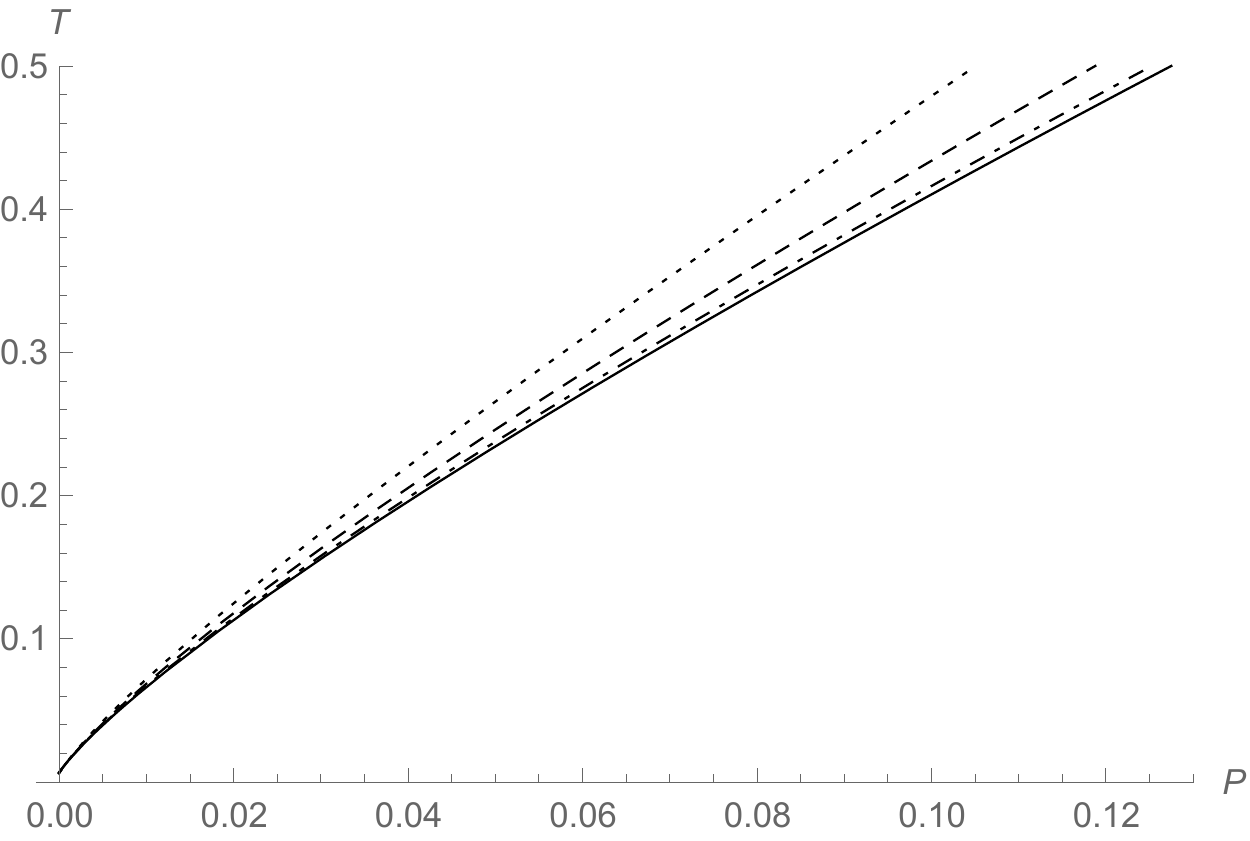}
	}
	\quad
	\subfigure[\quad $g=2,\,J=1,5,10,15$]
	{\centering
		\includegraphics[scale=0.57]{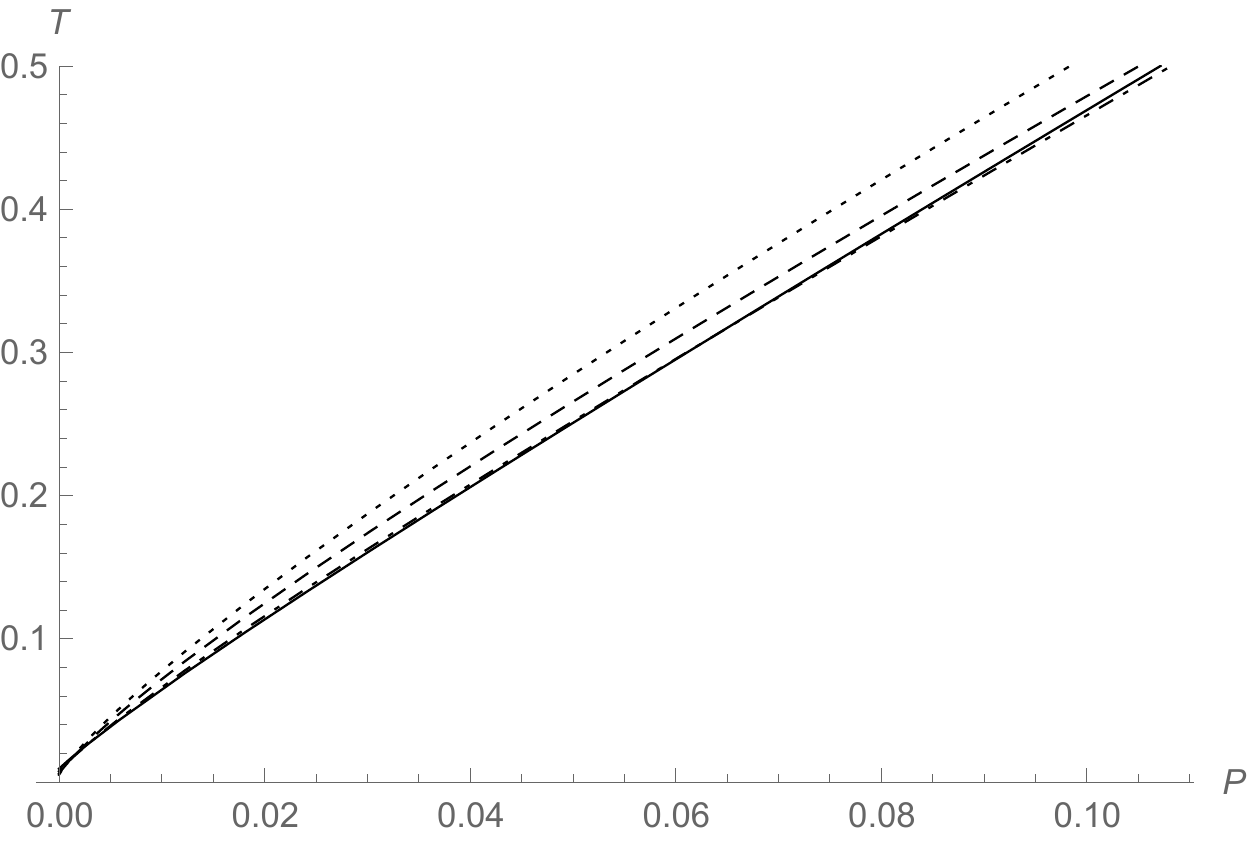}
	}
	\subfigure[\quad $j=15,\,g=0.5,1,1.5,2$]
	{\centering
		\includegraphics[scale=0.57]{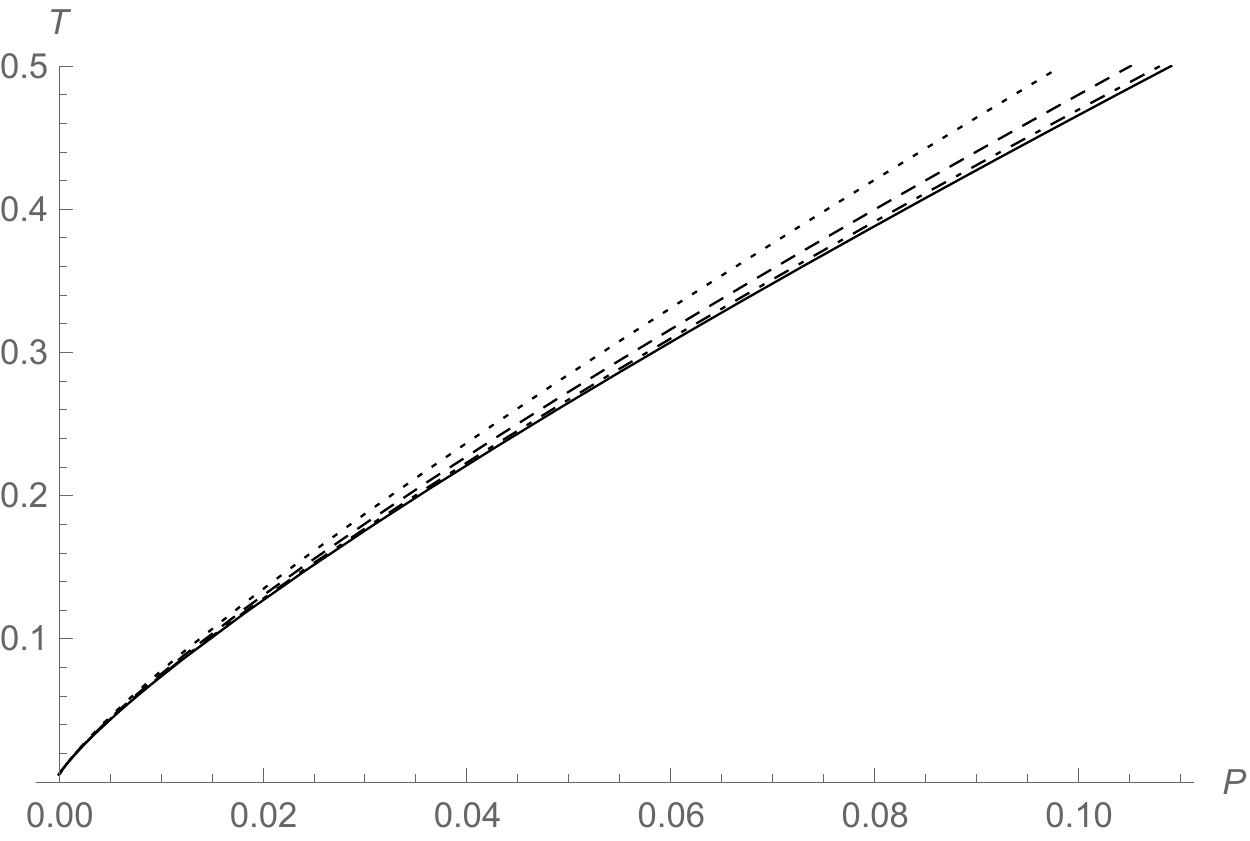}
	}
	\caption{Inversion curves of Rotating Bardeen AdS black holes. Each figure of the left column keeps $g$ fixed and with $J$ increasing from bottom to top,
		while each figure in the right column keeps $J$ fixed and with $g$ increasing from bottom to top.}
	\label{inversioncurve}
\end{figure*}

From Fig.\ref{inversioncurve}, we can observe that the inversion curves are always with positive slope like many cases studied before, hence, we only have the minimum inversion temperature $T_{min}$ but no maximum one, which distincts from the VdW gas case, so does it exists some kind of black holes that the inversion curves of it exhibits analogous behaviour to the case of VdW gas is still an open question. One can note that the slope of the curves increase with $g$ and $J$, and in high pressure, $T_i$ increase with $g$ and $J$, while in low pressure, it decrease with $g$ and $J$.

Moerover, there is another way to derive the $T_i$ which was first proposed in Ref.\cite{Zou:2013owa}, they used a mathematical trick, one can calculate the following equation
\begin{equation}
\left(\frac{{\partial (V/T)}}{{\partial S}}\right)_{P,J,g}=0,
\end{equation}
the result is in accordance with setting the numerator of $\mu$ to be zero, this method is much more convenient than calculate the JT coefficient then set it to be zero, because the mass $M$ is eliminated in $V/T$, which simplify the calculation greatly.

The second question, in order to get the minimum inversion temperature $T_{min}$, we set the pressure $P=0$, the numerator of $\mu$ becomes,  
\begin{equation}
\begin{aligned}
&-4 \pi ^6 g^{12}-4 \pi ^2 S^4 \left(3 g^4+4 J^2\right)+3 \pi  g^2 S^5\\
&-2 \pi ^4 S^2 \left(21 g^8+18 g^4 J^2+8 J^4\right)-\pi ^5 g^6 S \left(21 g^4+16 J^2\right)\\
&-2 \pi ^3 g^2 S^3 \left(19 g^4+18 J^2\right)+2 S^6=0
\end{aligned}
\end{equation}
this is also a lengthy equation, we can only use numerical method again to get the values of $S$ with given values of $J$ and $g$, then put it in Eq.\ref{T} to obtain the minimum inversion temperature $T_{min}$, finally, we calculate the ratio between $T_{min}$ and $T_C$. However, the critical temperature of rotating Bardeen-AdS black hole had not been obtained in any literature before, here we use the conditions $({\partial_S \,T})=0$ and $({\partial_{S,S}^2 \,T})=0$~\cite{Wei:2014qwa} to calculate $T_C$, however, the above two equations are quite complicated, hence, we use numerical method one more time, with specific values of $J$ and $g$ one can get the values of $S$ and $P$, then subsititute it into Eq.\ref{T}, we obtain $T_C$, finally, the three quantities are shown in Tab.\ref{table1}, the case comes back to Kerr-AdS black holes when $g=0$.

\begin{table}[!htbp]
	
	\begin{ruledtabular}
		\caption{$T_{min}$, $T_C$ and their ratio $T_{min}/\,T_C$ with $J=1$ but different values of $g$}
		\label{table1}
		\begin{spacing}{1.2}
		\begin{tabular}{lclclclclcl}
			
			&$g=0.5$&$g=1.0$&$g=1.5$&$g=2.0$ \\ \hline
			
			$T_C$&0.0387523&0.0304628&0.0224486&0.0172272 \\
			$T_{min}$&0.0197767&0.0159827&0.0119638&0.0092224\\
			$T_{min}/T_C$&0.510336&0.524663&0.532942&0.535339\\
			
		\end{tabular}
	\end{spacing}
	\end{ruledtabular}
\end{table}

From table.\ref{table1}, we can discover that with the nonlinear paramater $g$ included, there is a big difference from the former case, for the RN-AdS black holes, $T_{min}/\,T_C=1/2$~\cite{Okcu:2016tgt}; for the Kerr-AdS black holes, $T_{min}/\,T_C=0.504622$~\cite{Okcu:2017qgo}; for the Bardeen-AdS black holes, $T_{min}/\,T_C=0.5366622$~\cite{Pu:2019bxf}; for the KN-AdS black holes, $T_{min}/\,T_C$ lies between 0.4997 to 0.5096~\cite{Zhao:2018kpz}. The first three ratios are independent of other parameter, the last has slight dependence on the parameter $Q$ and $J$. However, in the rotatng Bardeen-AdS black holes case, we note that at certain values of $J$, the ratio increase with $g$, which is a little greater than $1/2$ but smaller than $1$, which means some kind of phase transition maybe happens during the throttling process. 

\begin{figure*}[htbp]
	\subfigure[\quad $J=1,\,g=0.5,\,M=2,2.5,3,3.5$]
	{\centering
		\includegraphics[scale=0.6]{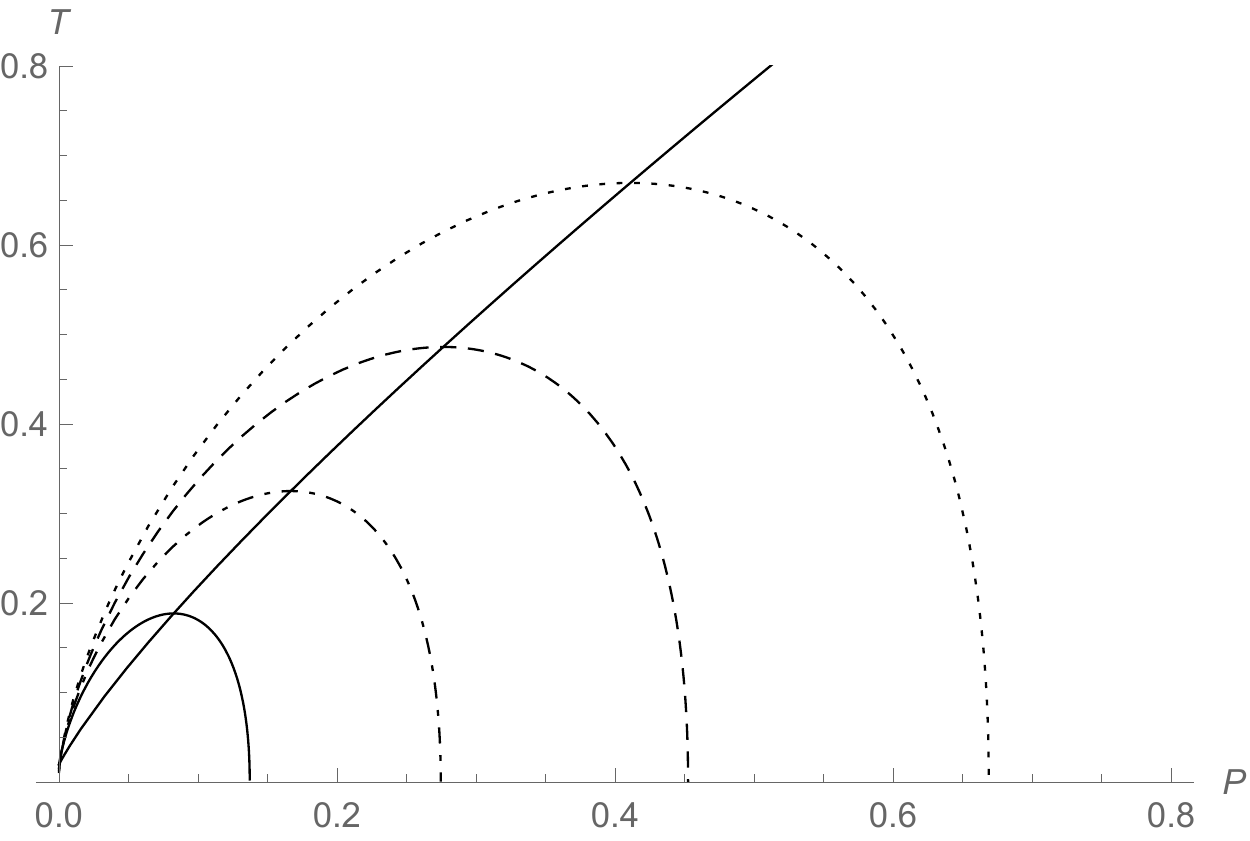}
	}	
	\subfigure[\quad $J=1,\,g=1,\,M=2,2.5,3,3.5$]
	{\centering
		\includegraphics[scale=0.6]{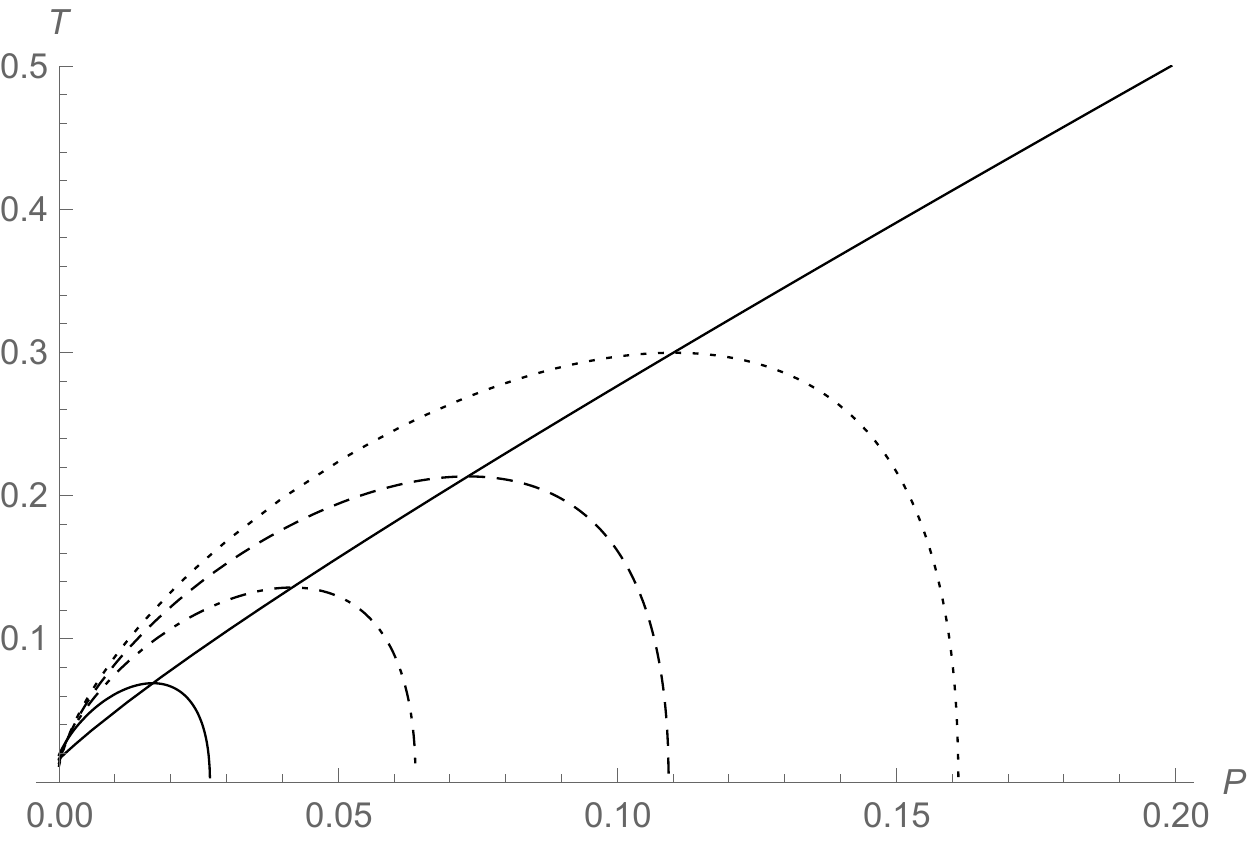}
	}
	\quad
	
	\subfigure[\quad $J=1,\,g=1.5,\,M=3,3.5,4,4.5$]
	{\centering
		\includegraphics[scale=0.6]{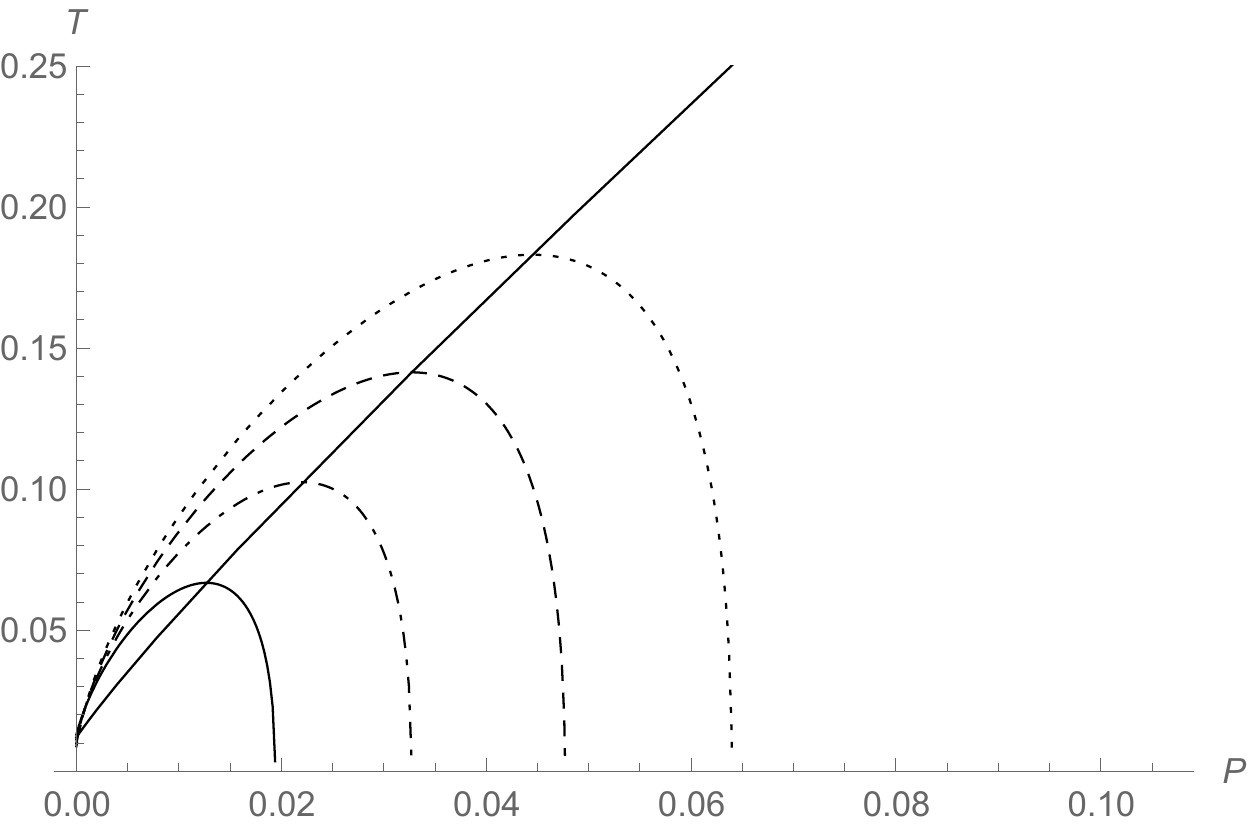}
	}
	\subfigure[\quad $J=1,\,g=2,\,M=5,5.5,6,6.5$]
	{\centering
		\includegraphics[scale=0.6]{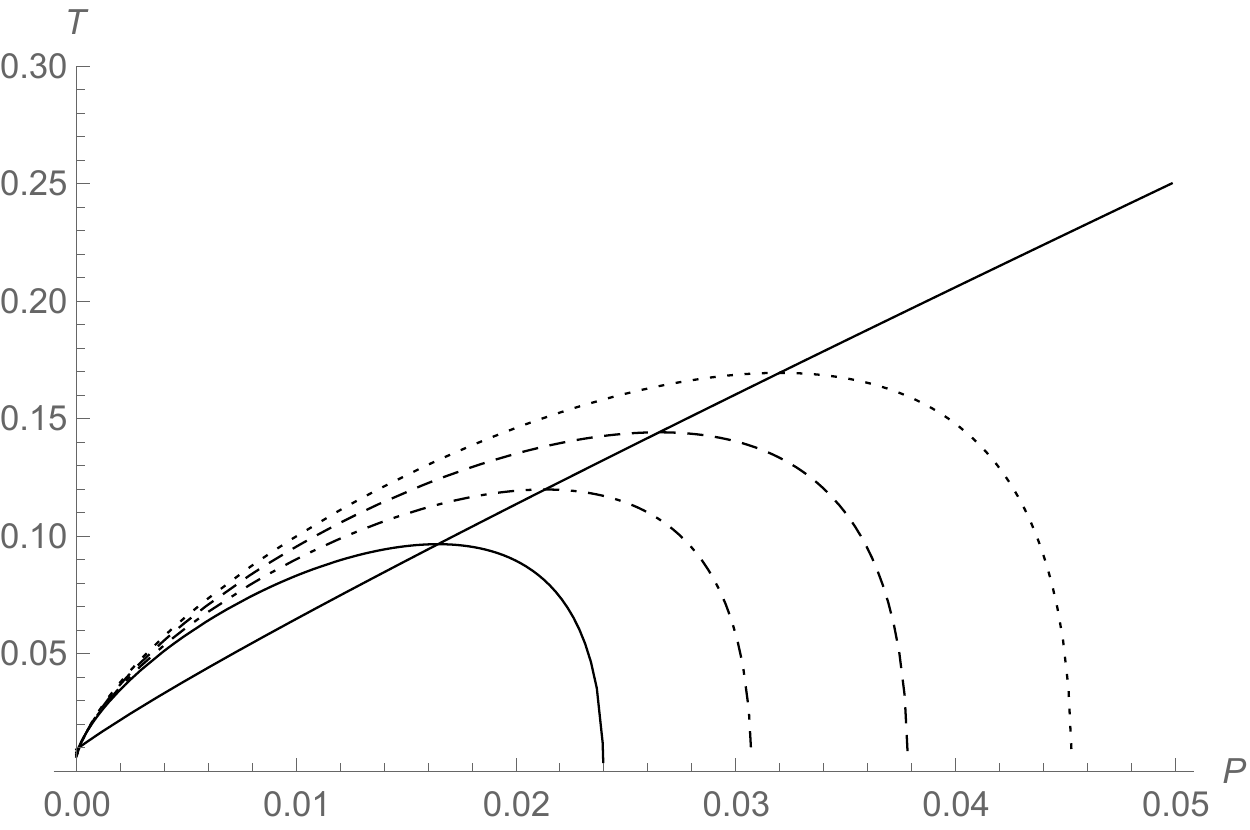}
	}
	\caption{Isenthalpic curves and inversion cirves of the Rotating Bardeen-AdS black holes. With $J$ fixed and the isenthalpic curves correspond to increasing values of $M$ from bottom to top.}
	\label{isenthalpic}
\end{figure*}

The third issue is to plot the isenthalpic curves in the $T-P$ plane, remember that the mass is equivalent to enthalpy in the extended phase space, thus we set the mass to be constant in Eq.\ref{M}, and with specific values of $g$ and $J$, we can plot the isenthalpic curves combined with Eq.\ref{T} by numerical method. The results are shown in Fig.\ref{isenthalpic}, as we can see, in each subfigure, the isenthalpic curves are seperated by the inversion curve into two parts, the intersection points are the extreme point of the isenthalpic curves, the JT coefficient in the left half part is positive which is called the cooling region, whlie in the right half part is negative which is called the heating region, besides, the entire shape of the isenthalpic are analogous to the previous cases.

\begin{table}[!htbp]
	\begin{ruledtabular}
		\caption{The minimum mass $M_{min}$ with different values of $g$ and $J$} 
		\label{table2}
		\begin{spacing}{1.2}
		\begin{tabular}{lclclclclcl}
			
			&$g=0.5$&$g=1.0$&$g=1.5$&$g=2.0$ \\ \hline
			$J=1$&1.14135&1.50015&2.06798&2.70698 \\
			$J=5$&2.36765&2.50411&2.75472&3.13398\\
			$J=10$&3.31786&3.41081&3.57509&3.82218\\
			$J=15$&4.05121&4.12612&4.25619&4.44859\\
		\end{tabular}
		\end{spacing}
	\end{ruledtabular}
\end{table}

\begin{figure}[htbp]
\centering
\includegraphics[scale=0.5]{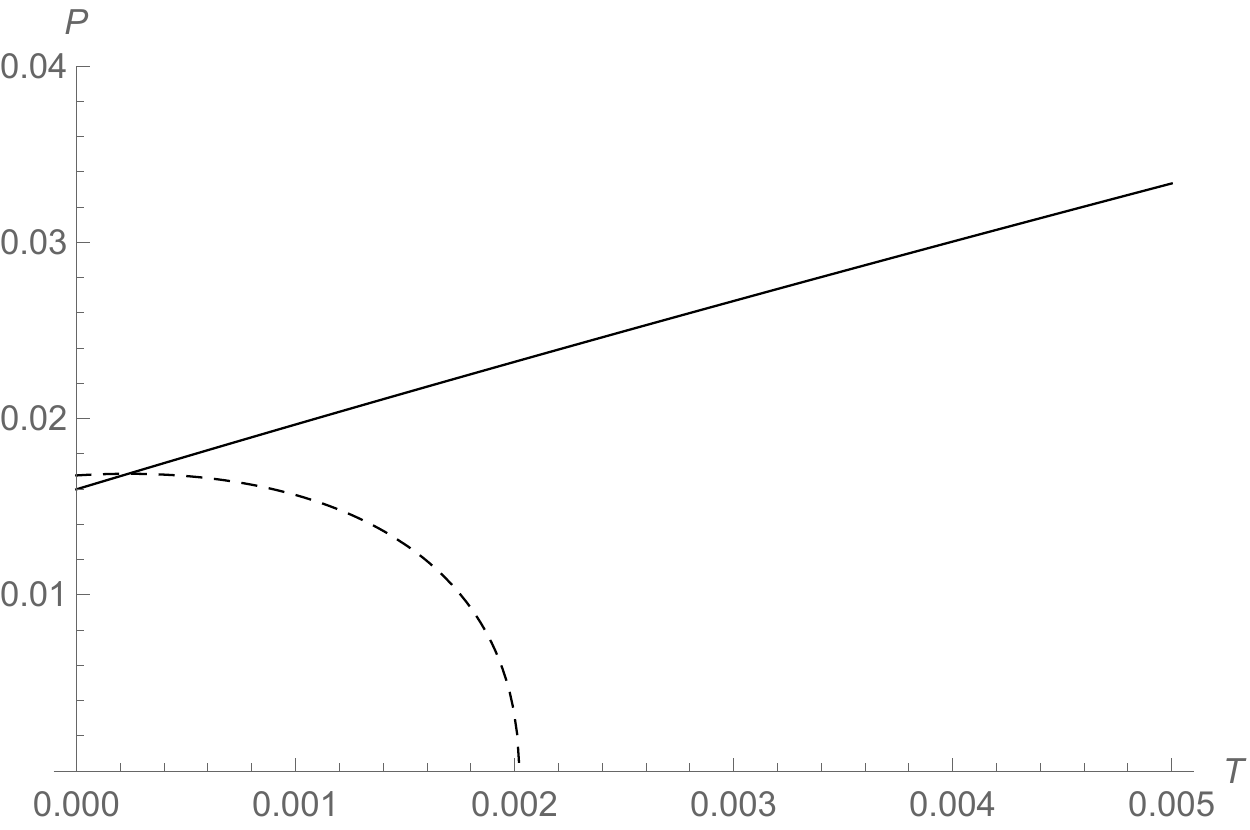}
\caption{Isenthalpic curve and the inversion curve when $g=1,\,J=1,\,M=1.51$}
\label{isenthalpicused}
\end{figure}

Finally, another interesting quantity was first discussed in Ref.\cite{Zou:2013owa}, it was shown that at given parameter values, if the mass is less than a certain values, then the isenthalpic curve has no intersection point with the inversion curve, that is to say, the black hole always in heating process. Three case have been computed in Ref.\cite{Zou:2013owa}, for the RN-AdS black holes, $M_{min}=5\sqrt{6}\,Q/12$; for the Kerr-AdS black holes, $M_{min}=\sqrt[4]{(2+3\sqrt{6})/8}\,\sqrt{J}$; for the KN-AdS black holes, $M_{min}=1.309$(when $J=1$ and $Q=1$). It is intriguing to investigate this issue in current situation, the minimum mass $M_{min}$ can immediately obtain as soon as we substitute the entropy $S$ corresponding to $T_{min}$ into Eq.\ref{M}, the results are shown in Tab.\ref{table2} with different values of $g$ and $J$. As we can see, $M_{min}$ increase with $g$ and $J$, this implies that the nonlinear parameter $g$ may exhibit some parallel effects as angular momentum $J$. For demonstration, in Fig.~\ref{isenthalpicused} we drawn the isenthalpic curve and inversion curve when $g=1$ and $J=1$, from Tab.~\ref{table2} we know in this case $M_{min}=1.50015$, and in Fig.~\ref{isenthalpicused}, the two curves are nearly separated, this confirm the numerical results above. 

\section{CONCLUSION}\label{con}
The throttling process of Rotating Bardeen-AdS black hole has been investigated in detail in this paper. In the extended phase space, the cosmological constant is treated as thermodynamic pressure and the mass of black hole as enthalpy, thus there are more fascinating thermodynamic properties in the extended phase space.

Firstly, we review the metric of the black hole, the necessary thermodynamic quantities are derived. Then the explicit JT coefficient is calculated via the expressions of temperature and volume, the curves of the JT coefficient as a function of entropy are also presented with different values of nonlinear parameter $g$ but with $J$ and $P$ fixed. 

Secondly, even though it is impossible to calculate the explicit form of the inversion temperature, yet the inversion curves and isenthalpic curves are drawn with numerical method, we notice that with increase in $g$ and $J$, the inversion curves tend to coincide with each other, but it seems that the nonlinear parameter $g$ has greater effect than the angular momentum $J$. And the isenthalpic curves are divided into two parts by the inversion curves, as long as the mass is no less than the minimum mass $M_{min}$, otherwise, the black holes are always in heating process, which means that with decrease of pressure the temperature increase. Finally, we work out the ratio between $T_{min}$ and $T_C$ by numerical method, and the minimum mass $M_{min}$ are obtained with ease, the ratio $T_{min}/\,T_{C}$ is dependent on $g$, and increase with it, which has not been found similar behaviour before.

To the end, we conclude that the nonlinear parameter $g$ has a nonnegligible effect on the thermodynamic properties of black hole, the thermodynamic properties black hole maybe exhibit some difference in the absence of singularity, more interesting phenomena are remained to be probed.

\begin{acknowledgments}
	The author expresses his sincere gratitude to the Beijing Normal University.
\end{acknowledgments}


\begin{thebibliography}{lab}
	
	\bibitem{Bardeen:1973gs} 
	J.~M.~Bardeen, B.~Carter and S.~W.~Hawking,
	Commun.\ Math.\ Phys.\  {\bf 31}, 161 (1973).
	doi:10.1007/BF01645742
	
	\bibitem{Bekenstein:1973ur} 
	J.~D.~Bekenstein,
	Phys.\ Rev.\ D {\bf 7}, 2333 (1973).
	doi:10.1103/PhysRevD.7.2333
	
	\bibitem{Hawking:1974sw} 
	S.~W.~Hawking,
	Commun.\ Math.\ Phys.\  {\bf 43}, 199 (1975)
	Erratum: [Commun.\ Math.\ Phys.\  {\bf 46}, 206 (1976)].
	doi:10.1007/BF02345020, 10.1007/BF01608497
	
	\bibitem{Hawking:1982dh} 
	S.~W.~Hawking and D.~N.~Page,
	Commun.\ Math.\ Phys.\  {\bf 87}, 577 (1983).
	doi:10.1007/BF01208266
	
	\bibitem{Maldacena:1997re} 
	J.~M.~Maldacena,
	Int.\ J.\ Theor.\ Phys.\  {\bf 38}, 1113 (1999)
	[Adv.\ Theor.\ Math.\ Phys.\  {\bf 2}, 231 (1998)]
	doi:10.1023/A:1026654312961, 10.4310/ATMP.1998.v2.n2.a1
	[hep-th/9711200].
	
	\bibitem{Chamblin:1999tk} 
	A.~Chamblin, R.~Emparan, C.~V.~Johnson and R.~C.~Myers,
	Phys.\ Rev.\ D {\bf 60}, 064018 (1999)
	doi:10.1103/PhysRevD.60.064018
	[hep-th/9902170].
	
	\bibitem{Chamblin:1999hg} 
	A.~Chamblin, R.~Emparan, C.~V.~Johnson and R.~C.~Myers,
	Phys.\ Rev.\ D {\bf 60}, 104026 (1999)
	doi:10.1103/PhysRevD.60.104026
	[hep-th/9904197].
	
	\bibitem{Kastor:2009wy} 
	D.~Kastor, S.~Ray and J.~Traschen,
	Class.\ Quant.\ Grav.\  {\bf 26}, 195011 (2009)
	doi:10.1088/0264-9381/26/19/195011
	[arXiv:0904.2765 [hep-th]].
	
	\bibitem{Dolan:2010ha} 
	B.~P.~Dolan,
	Class.\ Quant.\ Grav.\  {\bf 28}, 125020 (2011)
	doi:10.1088/0264-9381/28/12/125020
	[arXiv:1008.5023 [gr-qc]].
	
	\bibitem{Dolan:2011xt} 
	B.~P.~Dolan,
	Class.\ Quant.\ Grav.\  {\bf 28}, 235017 (2011)
	doi:10.1088/0264-9381/28/23/235017
	[arXiv:1106.6260 [gr-qc]].
	
	\bibitem{Kubiznak:2012wp} 
	D.~Kubiznak and R.~B.~Mann,
	JHEP {\bf 1207}, 033 (2012)
	doi:10.1007/JHEP07(2012)033
	[arXiv:1205.0559 [hep-th]].
	
	\bibitem{Gunasekaran:2012dq} 
	S.~Gunasekaran, R.~B.~Mann and D.~Kubiznak,
	JHEP {\bf 1211}, 110 (2012)
	doi:10.1007/JHEP11(2012)110
	[arXiv:1208.6251 [hep-th]].
	
	\bibitem{Zou:2013owa} 
	D.~C.~Zou, S.~J.~Zhang and B.~Wang,
	Phys.\ Rev.\ D {\bf 89}, no. 4, 044002 (2014)
	doi:10.1103/PhysRevD.89.044002
	[arXiv:1311.7299 [hep-th]].
	
	\bibitem{Cai:2013qga} 
	R.~G.~Cai, L.~M.~Cao, L.~Li and R.~Q.~Yang,
	JHEP {\bf 1309}, 005 (2013)
	doi:10.1007/JHEP09(2013)005
	[arXiv:1306.6233 [gr-qc]].
	
	\bibitem{Cheng:2016bpx} 
	P.~Cheng, S.~W.~Wei and Y.~X.~Liu,
	Phys.\ Rev.\ D {\bf 94}, 024025 (2016)
	doi:10.1103/PhysRevD.94.024025
	[arXiv:1603.08694 [gr-qc]].
	
	\bibitem{Okcu:2016tgt} 
	Ö.~Ökcü and E.~Aydıner,
	Eur.\ Phys.\ J.\ C {\bf 77}, no. 1, 24 (2017)
	doi:10.1140/epjc/s10052-017-4598-y
	[arXiv:1611.06327 [gr-qc]].
	
	\bibitem{Okcu:2017qgo} 
	Ö.~Ökcü and E.~Aydıner,
	Eur.\ Phys.\ J.\ C {\bf 78}, no. 2, 123 (2018)
	doi:10.1140/epjc/s10052-018-5602-x
	[arXiv:1709.06426 [gr-qc]].
	
	\bibitem{Zhao:2018kpz} 
	Z.~W.~Zhao, Y.~H.~Xiu and N.~Li,
	Phys.\ Rev.\ D {\bf 98}, no. 12, 124003 (2018)
	doi:10.1103/PhysRevD.98.124003
	[arXiv:1805.04861 [gr-qc]].
	
	\bibitem{Lan:2018nnp} 
	S.~Q.~Lan,
	Phys.\ Rev.\ D {\bf 98}, no. 8, 084014 (2018)
	doi:10.1103/PhysRevD.98.084014
	[arXiv:1805.05817 [gr-qc]].
	
	\bibitem{Mo:2018rgq} 
	J.~X.~Mo, G.~Q.~Li, S.~Q.~Lan and X.~B.~Xu,
	Phys.\ Rev.\ D {\bf 98}, no. 12, 124032 (2018)
	doi:10.1103/PhysRevD.98.124032
	[arXiv:1804.02650 [gr-qc]].
	
	\bibitem{Mo:2018qkt} 
	J.~X.~Mo and G.~Q.~Li,
	Class.\ Quant.\ Grav.\  {\bf 37}, no. 4, 045009 (2020)
	doi:10.1088/1361-6382/ab60b9
	[arXiv:1805.04327 [gr-qc]].
	
	\bibitem{Ghaffarnejad:2018exz} 
	H.~Ghaffarnejad, E.~Yaraie and M.~Farsam,
	Int.\ J.\ Theor.\ Phys.\  {\bf 57}, no. 6, 1671 (2018)
	doi:10.1007/s10773-018-3693-7
	[arXiv:1802.08749 [gr-qc]].
	
	\bibitem{Chabab:2018zix} 
	M.~Chabab, H.~El Moumni, S.~Iraoui, K.~Masmar and S.~Zhizeh,
	LHEP {\bf 02}, 05 (2018)
	doi:10.31526/LHEP.2.2018.02
	[arXiv:1804.10042 [gr-qc]].
	
	\bibitem{Cisterna:2018jqg} 
	A.~Cisterna, S.~Q.~Hu and X.~M.~Kuang,
	Phys.\ Lett.\ B {\bf 797}, 134883 (2019)
	doi:10.1016/j.physletb.2019.134883
	[arXiv:1808.07392 [gr-qc]].
	
	\bibitem{AhmedRizwan:2019yxk} 
	A.~Rizwan C.L., N.~Kumara A., D.~Vaid and K.~M.~Ajith,
	Int.\ J.\ Mod.\ Phys.\ A {\bf 33}, no. 35, 1850210 (2019)
	doi:10.1142/S0217751X1850210X
	[arXiv:1805.11053 [gr-qc]].
	
	\bibitem{Yekta:2019wmt} 
	D.~Mahdavian Yekta, A.~Hadikhani and Ö.~Ökcü,
	Phys.\ Lett.\ B {\bf 795}, 521 (2019)
	doi:10.1016/j.physletb.2019.06.049
	[arXiv:1905.03057 [hep-th]].
	
	\bibitem{Guo:2019gkr} 
	S.~Guo, J.~Pu and Q.~Q.~Jiang,
	arXiv:1905.03604 [gr-qc].
	
	\bibitem{Li:2019jcd} 
	C.~Li, P.~He, P.~Li and J.~B.~Deng,
	arXiv:1904.09548 [gr-qc].
	
	\bibitem{Pu:2019bxf} 
	J.~Pu, S.~Guo, Q.~Q.~Jiang and X.~T.~Zu,
	arXiv:1905.02318 [gr-qc].
	
	\bibitem{Rostami:2019ivr} 
	M.~Rostami, J.~Sadeghi, S.~Miraboutalebi, A.~A.~Masoudi and B.~Pourhassan,
	arXiv:1908.08410 [gr-qc].
	
	\bibitem{Haldar:2018cks} 
	A.~Haldar and R.~Biswas,
	EPL {\bf 123}, no. 4, 40005 (2018).
	doi:10.1209/0295-5075/123/40005
	
	\bibitem{Sadeghi:2020bon} 
	J.~Sadeghi and R.~Toorandaz,
	Nucl.\ Phys.\ B {\bf 951}, 114902 (2020).
	doi:10.1016/j.nuclphysb.2019.114902
	
	\bibitem{Guo:2019pzq} 
	S.~Guo, Y.~Han and G.~P.~Li,
	arXiv:1912.09590 [hep-th].
	
	\bibitem{Lan:2019kak} 
	S.~Q.~Lan,
	Nucl.\ Phys.\ B {\bf 948}, 114787 (2019).
	doi:10.1016/j.nuclphysb.2019.114787
	
	\bibitem{Nam:2020gud} 
	C.~H.~Nam,
	Eur.\ Phys.\ J.\ Plus {\bf 135}, no. 2, 259 (2020).
	doi:10.1140/epjp/s13360-020-00274-2
	
	\bibitem{K.:2020rzl} 
	R.~K., C.~L.~A.~Rizwan, A.~N.~Kumara, D.~Vaid and M.~S.~Ali,
	arXiv:2002.03634 [gr-qc].
	
	\bibitem{Hawking:1973uf} 
	S.~W.~Hawking and G.~F.~R.~Ellis,
	doi:10.1017/CBO9780511524646
	
	\bibitem{bar} 
	J.~M.~Bardeen,
	in {\it Conference Proceedings of GR5} (Tbilisi, USSR, 1968), p. 174. 
	
	\bibitem{AyonBeato:2000zs} 
	E.~Ayon-Beato and A.~Garcia,
	Phys.\ Lett.\ B {\bf 493}, 149 (2000)
	doi:10.1016/S0370-2693(00)01125-4
	[gr-qc/0009077].
	
	\bibitem{AyonBeato:1999rg} 
	E.~Ayon-Beato and A.~Garcia,
	Phys.\ Lett.\ B {\bf 464}, 25 (1999)
	doi:10.1016/S0370-2693(99)01038-2
	[hep-th/9911174].
	
	\bibitem{Hayward:2005gi} 
	S.~A.~Hayward,
	Phys.\ Rev.\ Lett.\  {\bf 96}, 031103 (2006)
	doi:10.1103/PhysRevLett.96.031103
	[gr-qc/0506126].
	
	\bibitem{Bambi:2013ufa} 
	C.~Bambi and L.~Modesto,
	Phys.\ Lett.\ B {\bf 721}, 329 (2013)
	doi:10.1016/j.physletb.2013.03.025
	[arXiv:1302.6075 [gr-qc]].
	
	\bibitem{Ali:2019myr} 
	M.~S.~Ali and S.~G.~Ghosh,
	Phys.\ Rev.\ D {\bf 99}, no. 2, 024015 (2019).
	doi:10.1103/PhysRevD.99.024015
	
	\bibitem{Caldarelli:1999xj} 
	M.~M.~Caldarelli, G.~Cognola and D.~Klemm,
	Class.\ Quant.\ Grav.\  {\bf 17}, 399 (2000)
	doi:10.1088/0264-9381/17/2/310
	[hep-th/9908022].
	
	\bibitem{Wei:2014qwa} 
	S.~W.~Wei and Y.~X.~Liu,
	Phys.\ Rev.\ D {\bf 91}, no. 4, 044018 (2015)
	doi:10.1103/PhysRevD.91.044018
	[arXiv:1411.5749 [hep-th]].
	
	
	
\end{thebibliography}
\end{document}